\documentclass[11pt,a4paper]{article}
\usepackage{authblk}
\usepackage{graphicx} 
\usepackage{amsfonts}
\usepackage{braket}

\title{Computing the molecular ground state energy in a restricted active space using quantum annealing}
\author[1]{Stefano Bruni}
\affil[1]{Department of Physics Aldo Pontremoli,Università degli Studi di Milano, Via Celoria 16, 20133 Milano, Italy}
\author[1]{Enrico Prati \thanks{enrico.prati@unimi.it}}

\date{December 2025}
\usepackage[backend=biber,style=numeric,sorting=none]{biblatex}
\usepackage{amsmath} 
\DeclareUnicodeCharacter{A3AC}{}
\begin{document}

\maketitle
\begin{abstract}
Calculating the molecular ground-state energy is a central challenge in computational chemistry. Conventional methods such as the Complete Active Space Configuration Interaction scale exponentially with molecular size, limiting their applicability to large molecules. 
Quantum computing offers a promising alternative by mapping molecular Hamiltonians by qubits, enabling cheaper computational scaling. Previous studies have shown that it is possible to formulate molecular ground state calculations as discrete optimization problems, addressable by quantum annealing. However, these efforts have been limited by previous generations of hardware and suboptimal annealing techniques.
Here, the H$_2$O ground-state problem is mapped to an Ising Hamiltonian using the Xian–Bias–Kas (XBK) method. By taking advantage of enhanced qubit connectivity and shorter embedding chains, it is solved with a more than doubled probability of achieving Hartree–Fock–level solutions with respect to the most advanced predecessor. Advanced annealing strategies extend Hartree–Fock–level accuracy to significantly larger problem instances, enabling solutions that use nearly 2.5 times more physically embedded qubits than the largest cases previously reported and allowing to improve annealing results by two orders of magnitude, reaching an energy difference of 0.120~Hartree relative to Hartree-Fock. These results show tangible progress toward practical quantum annealing applications in NISQ era.
\end{abstract}

\section{Introduction}
Accurate calculation of molecular ground-state energy is a fundamental challenge in quantum chemistry, enabling precise predictions of molecular stability and electronic correlations. Their knowledge is essential for the design of catalysts \cite{catalysts}, pharmaceuticals \cite{pharmaceutical}, and quantum materials \cite{materials}, just to mention a few. Despite advances in classical electronic-structure methods, from post-Hartree–Fock \cite{postHF} to density-functional approaches \cite{DFT}, achieving chemical accuracy still requires computational resources that scale exponentially with system size. Consequently, these methods are impractical for molecules beyond a few tens of atoms\cite{Limits}, far smaller than many chemically or industrially relevant systems. Quantum computing offers a fundamentally different paradigm. It consists of encoding and manipulating quantum states directly within a Hilbert space of dimension $2^n$. Quantum processors can, in principle, capture electronic correlations with polynomial resources, offering a method to simulate ground-state properties beyond classical limits\cite{QC}.
The development of such quantum computing approach becomes therefore essential for advancing the accurate modeling of molecules and design of materials of practical and industrial significance\cite{QCforMolecules}.

Classical electronic-structure methods, such as Hartree–Fock (HF) \cite{HF} and Complete Active Space Configuration Interaction (CASCI)\cite{CASCI}, provide the foundation for molecular ground state calculations. HF offers a mean-field approximation that is computationally efficient but neglects electron correlation beyond the averaged interaction, while CASCI includes explicit electron correlation and can therefore achieve chemical accuracy for small molecules. Such methods face fundamental scaling challenges. Indeed, HF calculations scale roughly as $O(N^4)$ with the number of basis functions\cite{HFScaling}, while CASCI grows exponentially with the number of active electrons and orbitals, severely limiting the ability to model complex molecules. To overcome the above limitations, a variety of classical approaches have been explored, including truncated configuration interaction, perturbative corrections, and approximate density functional methods \cite{QChemistryOverview}. While the latter strategies extend the reach of classical calculations, they often involve trade-offs between accuracy and computational cost, leaving large or strongly correlated systems inaccessible.

Quantum computing promises significant improvements in quantum chemistry, offering better scaling for correlated electronic structure methods. Among current platforms, D-Wave quantum annealers stand out for their stability and number of qubits amounting to 4593 units in the latest release. Quantum annealers leverage energy landscape exploration via quantum phenomena, promising a more efficient search over combinatorial configurations of quantum states than classical heuristics \cite{QA}. Previous studies have demonstrated the feasibility of applying these devices to molecular ground state calculations by translating CASCI problems into an Ising formulation using the Xian-Bias-Kas (XBK) method \cite{copenhaver}. Benchmarks on small molecules such as H$_2$O have shown controlled accuracy improvements and confirmed that quantum annealing can effectively encode electronic correlations. However, the potential of the latest Advantage 2 architecture and advanced annealing protocols, such as reverse annealing or paused schedules, remains largely unexplored, leaving open the question of how far current quantum annealers can advance practical quantum chemistry applications.
Some of us already applied quantum annealing to gravity inversion problems\cite{siddi}, optimal antenna placement \cite{Nigro} and Boltzmann machines\cite{rocutto2021complete,moro,noe}, including the exploitation of reverse annealing\cite{rocutto}.
Here, the representative H$_2$O ground state problem is mapped onto an Ising Hamiltonian using the Xian-Bias-Kas (XBK) method \cite{XBK} and solved on the latest quantum annealing processor of the D-Wave Company. The D-Wave hyperparameters are chosen to ensure sufficiently strong embedding chains while minimizing accuracy loss caused by hardware limitations in the Ising representation. It then follows a comprehensive study of how the number of required physical qubits and the QUBO energy gap scale as the method's accuracy is increased, in order to capture electron correlation, establishing the computational limitations of current annealing hardware and addressing a gap that has not been explored in previous literature. The new architecture is then benchmarked against its predecessor, Advantage 1, revealing substantial improvements in effective qubit connectivity, reduced embedding chain lengths, higher probability of obtaining Hartree–Fock–level solutions (130\% increase), and significantly lower QPU readout times, reduced by around 20\%. Additionally, by employing advanced annealing protocols such as reverse annealing and paused search, the accuracy achieved with standard schedules is improved by nearly two orders of magnitude for qubit instances with 50\% more logical qubits and nearly 150\% more physical qubits, which previous hardware was not able to resolve.
\begin{figure}[h]
    \centering
    \includegraphics[width=0.9\linewidth, clip]{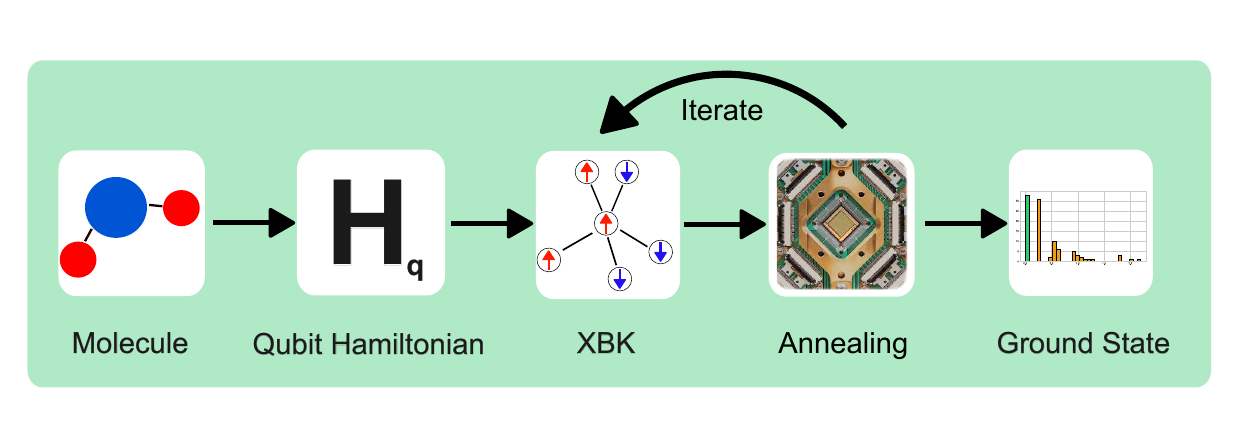}
    \caption{Workflow behind the ground state calculation using the XBK Quantum Annealer pipeline.}
    \label{fig:2}
\end{figure}
\section{Background of quantum annealing}
\subsection{Embodiment of Hamiltonian dynamics}
Quantum annealing (QA) is an heuristic optimization technique that exploits quantum fluctuations to search for the ground state of a problem Hamiltonian \cite{QAMath}. It is designed to solve problems that can be formulated as finding the global minimum of a cost function, which is encoded into a target Hamiltonian \( H_t \). The method relies on the adiabatic evolution of a quantum system governed by a time-dependent Hamiltonian of the form
\begin{equation}
    H(t) = A(t) H_i + B(t) H_t,
\end{equation}
where \( H_i \) is a driver Hamiltonian, \( H_t \) encodes the optimization problem, and \( A(t) \) and \( B(t) \) are monotonic annealing schedules satisfying \( A(0) \gg B(0) \) and \( A(T) \ll B(T) \) over the total annealing time \( T \).
The system is initially prepared in the easily accessible ground state of \( H_i \), and under sufficiently slow evolution, the adiabatic theorem \cite{adiabatic} guarantees convergence towards the ground state of \( H_t \).

In practical implementations, such as those used in D-Wave quantum annealers, the target Hamiltonian takes the form of the Ising model:
\begin{equation}
    H_I = \sum_i h_i \sigma_i^z + \sum_{i<j} J_{ij} \sigma_i^z \sigma_j^z,
\end{equation}
where \( h_i \) are local fields, \( J_{ij} \) are pairwise couplings, and \( \sigma_i^z \) are Pauli operators acting on spin-\(\tfrac{1}{2}\) qubits. 
Classical quadratic unconstrained binary optimization (QUBO) problems \cite{QUBO} can be mapped to this Ising form by replacing binary variables \( x_i \in \{0,1\} \) with spin variables \( s_i = 2x_i - 1 \in \{-1,1\} \). 
Thus, any discrete optimization problem expressible as a quadratic cost function can, in principle, be implemented on a quantum annealer by appropriately setting \(\{h_i, J_{ij}\}\).

\subsection{Embedding of quantum annealing on D-Wave hardware}

D-Wave quantum annealers implement the time-dependent Hamiltonian
\begin{equation}
    H(t) = A(t) \sum_i \sigma_i^x + B(t) \left( \sum_i h_i \sigma_i^z + \sum_{i<j} J_{ij} \sigma_i^z \sigma_j^z \right),
\end{equation}
where the transverse-field term \( \sum_i \sigma_i^x \) acts as the driver Hamiltonian \( H_i \), introducing quantum fluctuations that enable tunneling between classical minima. 
In principle, the Ising model assumes all-to-all connectivity among qubits. 
However, D-Wave’s hardware implements a restricted physical connectivity graph, meaning that not all qubits can interact directly. 
This creates a distinction between the \textit{logical} Ising Hamiltonian, defined by the problem’s ideal couplings, and the \textit{physical} Hamiltonian that can be realized on the chip. 
To resolve this mismatch, D-Wave employs an \textit{embedding} procedure that maps each logical qubit onto a chain of physically connected qubits. 
Strong ferromagnetic couplings are used to align the qubits within a chain so that they collectively behave as a single logical unit.

As the logical problem becomes denser or larger, longer chains are required, which increases the probability of \textit{chain breaks}, instances where qubits within a chain fail to maintain alignment, thus reducing solution fidelity. 
Empirically, chain lengths above 8 are often found to be critical for maintaining reliable embeddings. Experiments reveal a clear correlation between the average chain length of embeddings and the relative errors of the solutions sampled \cite{embeddingandchains}.

The most recent D-Wave architectures, \textit{Pegasus} and \textit{Zephyr}, differ mainly in their connectivity density and embedding efficiency. 
Zephyr, implemented in the Advantage~2 system, increases the average qubit degree (20 couplers per qubit) and reduces embedding overhead, allowing larger and more connected problem graphs to be represented with shorter and therefore more stable chains. 

\section{Mapping a molecular Hamiltonian in Ising formulation}
This section outlines the steps required to express the molecular ground-state energy problem in an Ising formulation. The procedure begins by specifying the molecular geometry and selecting an appropriate basis set for the Hartree–Fock (HF) calculation. The Hamiltonian is then constructed in the second-quantised formalism using the HF orbitals, after which a fermion-to-qubit encoding is applied to obtain its representation in terms of Pauli operators. Next, qubit tapering techniques are used to reduce the qubit count. The XBK algorithm is subsequently employed to transform the qubit Hamiltonian into an Ising model, which is finally quadratised so that it only features quadratic (2-local) interactions. The pipeline corresponds to the first three steps of the diagram reported in Figure~\ref{fig:2}. 

\subsection{From Molecular specification to second-quantized Hamiltonian}

The first step in constructing a molecular Hamiltonian suitable for quantum computation is the definition of the molecular system. 
The geometry, charge, and spin multiplicity of the molecule are specified and provided to \texttt{OpenFermion} \cite{openfermion}, which serves as the primary electronic-structure interface. 
A basis set must also be chosen to represent the molecular orbitals; common choices include minimal and split-valence basis sets such as STO-3G, 6-31G, or the more accurate correlation-consistent sets (e.g., cc-pVDZ) \cite{basissets}. 
Larger basis sets improve the accuracy of the electronic description at the cost of increased computational complexity and a larger number of resulting qubits.

Once the molecular configuration and basis set are defined, \texttt{OpenFermion} performs a Hartree–Fock (HF) calculation to obtain the molecular orbitals. 
These orbitals are then used to compute the one- and two-electron integrals that define the molecular electronic Hamiltonian in its second-quantized form:
\begin{equation}
    H = \sum_{pq} h_{pq} a_p^\dagger a_q + \frac{1}{2} \sum_{pqrs} h_{pqrs} a_p^\dagger a_q^\dagger a_r a_s,
\end{equation}
where \( a_p^\dagger \) and \( a_q \) are fermionic creation and annihilation operators acting on spin orbitals, and \( h_{pq} \) and \( h_{pqrs} \) are the one- and two-electron integrals derived from the HF molecular orbitals. 
This Hamiltonian fully encodes the electronic interactions within the chosen active space and forms the starting point for mapping the molecular problem onto a qubit representation.
Since the number of spin orbitals directly determines the number of qubits required to encode the system, practical implementations typically employ a \textit{restricted active space} approach \cite{RAS}, where only the most relevant orbitals are retained. 
For instance, in the case of the H$_2$O molecule, computations can be restricted to the four highest occupied and lowest unoccupied molecular orbitals, corresponding to eight spin orbitals and four active electrons \cite{copenhaver}. This approximation significantly reduces the qubit requirements while preserving the essential electronic structure needed to describe the ground state.

\subsection{Pauli Hamiltonian and Fermion-to-Qubit Encodings}

Quantum computers operate on qubits rather than fermionic degrees of freedom, and therefore require that the second-quantized molecular Hamiltonian be expressed in terms of qubit operators.
This transformation, known as \textit{fermion-to-qubit encoding}, replaces the fermionic creation and annihilation operators \( a_p^\dagger, a_p \) with tensor products of Pauli matrices \( X, Y, Z \) acting on qubits. 
The mapping is non-trivial since fermionic creation and annihilation operators obey canonical anti-commutation relations, whereas Pauli operators form a non-commuting algebra. The encoding must therefore reproduce the fermionic statistics within a qubit-based representation.
A common and conceptually straightforward encoding is the \textit{Jordan–Wigner transformation} \cite{JW}, which represents the fermionic operators as:
\begin{equation}
    a_p^\dagger = \frac{1}{2} (X_p - iY_p) \prod_{j=0}^{p-1} Z_j, 
    \quad
    a_p = \frac{1}{2} (X_p + iY_p) \prod_{j=0}^{p-1} Z_j.
\end{equation}
The string of \( Z \) operators enforces the correct fermionic anti-commutation relations by keeping track of the particle parity for all lower-indexed orbitals. 
Although the Jordan–Wigner mapping is conceptually simple, it introduces non-local qubit interactions that scale linearly with the number of orbitals, which can limit its efficiency on hardware with restricted qubit connectivity.

Alternative mappings have been developed to reduce operator length and improve performance on near-term devices. 
The \textit{Bravyi–Kitaev transformation} \cite{BK}optimizes the balance between local occupation and parity information, yielding shorter Pauli strings on average. 
More recent encodings, such as \textit{Parity} mapping, further exploit system symmetries (e.g., particle number and spin conservation) to reduce the effective number of qubits required to represent the molecular Hamiltonian. 
Once this encoding is performed, the resulting Hamiltonian can be expressed as a weighted sum of Pauli strings,
\begin{equation}
    H = \sum_i c_i P_i, \quad P_i \in \{I, X, Y, Z\}^{\otimes n},
\end{equation}
where the coefficients \( c_i \) are real and \( n \) is the number of qubits in the active space.
These transformations can be performed automatically using \texttt{OpenFermion}, and in this work Parity Encoding is utilized.

\subsection{Qubit tapering and symmetry exploitation}

In current quantum hardware, reducing the number of qubits required for simulation is essential. Qubit tapering achieves this by exploiting symmetries in the molecular Hamiltonian \cite{tapering}. If a qubit is acted upon trivially or by the same Pauli operator in all terms, it can be removed by substituting the operator with its eigenvalue \(\pm 1\). More generally, let \(S\subset \mathbb{P}_M\) denote an abelian group of symmetry operators commuting with \(H\). By applying suitable Clifford transformations \(U_i\), these symmetries can be mapped to single-qubit operators \(\sigma_x^{(q(i))}\) such that
\begin{equation}
[U_i^\dagger H U_i, \sigma_x^{(q(i))}] = 0,
\end{equation}
allowing the corresponding qubits to be replaced with fixed eigenvalues and removed from the computation.

Symmetry generators can be efficiently identified using binary encodings of Pauli strings, which allow systematic determination of commuting operators \cite{PGSymmetries}. Once identified, the Hamiltonian is transformed via a sequence of Clifford operations and permutations:
\begin{equation}
H' = (U_1 W_1 \cdots U_k W_k) H (W_k^\dagger U_k^\dagger \cdots W_1^\dagger U_1^\dagger),
\end{equation}
producing a reduced Hamiltonian \(H'\) with fewer active qubits.

For typical molecular systems, such as H\(_2\)O in a minimal STO-3G basis, this procedure can halve the number of qubits required, significantly reducing the complexity of subsequent quantum computations. The process is fully automated in \texttt{OpenFermion} using \texttt{taper\_off\_qubits}.

\subsection{The XBK Method}
The qubit Hamiltonian obtained after tapering still contains Pauli operators in the \(X\) and \(Y\) directions, which cannot be directly implemented on a quantum annealer. The XBK method \cite{XBK} transforms the Hamiltonian by an expression compatible with annealers by expressing all operators using only \(Z\)-type Pauli matrices. It is achieved by introducing an expanded qubit space of dimension \(M' > M\), where the action of \(X\) and \(Y\) operators is encoded via additional qubits and auxiliary constraints.
Pauli operators are mapped as
\begin{equation}
\sigma_x^i \rightarrow \frac{1 - \sigma_z^{ij} \sigma_z^{ik}}{2}, \quad
\sigma_y^i \rightarrow i \frac{\sigma_z^{ik} - \sigma_z^{ij}}{2}, \quad
\sigma_z^i \rightarrow \frac{\sigma_z^{ij} + \sigma_z^{ik}}{2}, \quad
I^i \rightarrow \frac{1 + \sigma_z^{ij} \sigma_z^{ik}}{2},
\end{equation}
producing sub-Hamiltonians \(H^{(i,j)}\), that from now on will be referred to as XBK Hamiltonian Sectors, on a \(rm\)-qubit space. The \(rm\)-qubit Hamiltonians are constructed as
\begin{equation}
H'_p = \sum_{i,j \le r} H^{(i,j)} S_p(i) S_p(j),
\end{equation}
where \(S_p(i) = \pm 1\) accounts for sign combinations over \(\left\lfloor r/2 \right\rfloor\) sectors.  

To extract eigenvalues, one defines
\begin{equation}
D_{p,\lambda} = H'_p - \lambda C_p, \quad 
C_p = \sum_{\pm} \left[\sum_{i=1}^r \left(S_p(i) \prod_{k=1_i}^{m_i} \frac{1\pm \sigma_z^k}{2}\right) \right]^2,
\end{equation}
and iteratively adjusts \(\lambda\) until the minimum eigenvalue of \(D_{p,\lambda}\) is non-negative, yielding the ground-state energy \(\lambda'\). The corresponding \(m\)-qubit state coefficients can be approximated as
\begin{equation}
a_i \approx \frac{b_i S(b_i)}{\sqrt{\sum_j b_j^2}},
\end{equation}
with \(b_i\) the occurrence count of basis states and \(S(b_i)\) their combined signs.  

Before submission to the quantum annealer, \(D_{p,\lambda}\) must be quadratised to ensure all terms are at most 2-local, a step discussed in the following Section.

\textbf{Remark on the role of the \(r\) parameter.} The parameter \(r\) in the XBK method controls the number of replicas of the original qubits and directly impacts the precision of quantum chemical calculations. From now on we shall refer to the number of qubits in the expanded Hilbert space as \textit{XBK qubits}.
For \(r=1\), the XBK method represents a single Slater determinant, equivalent to the Hartree-Fock (HF) approximation, \(\Phi_{HF}\). Increasing \(r\) allows the method to represent a larger set of determinants, therefore capturing more correlation effects and approaching the accuracy of Complete Active Space Configuration Interaction (CASCI), which considers all configurations within the active space.

For \(r=1\), \(E_{HF} = \langle \Phi_{HF} | \hat{H} | \Phi_{HF} \rangle\), while for general \(r\), the XBK energy is
\begin{equation}
E_{XBK} = \sum_{i,j} c_i^* c_j \langle \Phi_i | \hat{H} | \Phi_j \rangle,
\end{equation}
with coefficients \(c_i\) increasing as \(r\) grows, approximating the CASCI energy
\begin{equation}
E_{CASCI} = \sum_{i,j} d_i^* d_j \langle \Phi_i | \hat{H} | \Phi_j \rangle,
\end{equation}
where \(d_i\) are determined from all active-space configurations.  

For small molecules like H\(_2\), \(r=16\) is sufficient to reach CASCI precision. However, such large \(r\) values rapidly become prohibitive for larger molecules, as the required number of qubits exceeds current hardware capabilities. This limitation is explored further in the context of H\(_2\)O in the next section.

\subsection{Quadratisation Procedure}
The Z-only Hamiltonian obtained via the XBK method may contain high-order terms up to the total number of qubits \(rm\). Such \(n\)-local interactions cannot be directly implemented on quantum annealers, which only support quadratic (2-local) Hamiltonians. Transforming a higher-order Hamiltonian into a quadratic form, known as \emph{quadratisation}, is therefore essential and can become computationally demanding for large systems.

Consider a simple three-body (3-local) term, \(x_1 x_2 x_3\), representing the simplest Higher-Order Binary Optimization (HUBO) problem. We introduce an ancillary variable \(y_1 := x_1 x_2\), allowing us to rewrite the term as \(y_1 x_3\), which is quadratic (QUBO) in form. To ensure that \(y_1\) satisfies the Boolean condition \(y_1 \iff x_1 \land x_2\), a penalty term is added:

\[
P(y_1, x_1, x_2) = x_1 x_2 - 2 (x_1 + x_2) y_1 + 3 y_1.
\]

The penalty term enforces the constraint on \(y_1\), guaranteeing that the minimum-energy solution respects the original 3-local relation. The final quadratised Hamiltonian then reads:

\[
H_\text{QUBO} = P(y_1, x_1, x_2) + y_1 x_3,
\]

which contains only quadratic terms and can be directly implemented on a quantum annealer. D-Wave provides automated quadratisation routines that perform this transformation and require a \emph{chain strength} parameter to enforce constraints reliably without introducing spurious energy minima.

With the quadratisation step completed, the Ising Hamiltonian is ready for hardware embedding, after which the QPU can be called to execute the annealing schedule. These final stages correspond to the last two blocks in Figure \ref{fig:2}. The annealing results are discussed in the next Section.

\section{Results}


\subsection{Ground state energy determination based on quantum annealing}

\begin{figure}  
    \centering
    \includegraphics[width=0.9\linewidth, , trim={0cm 6cm 0cm 0cm}, clip]{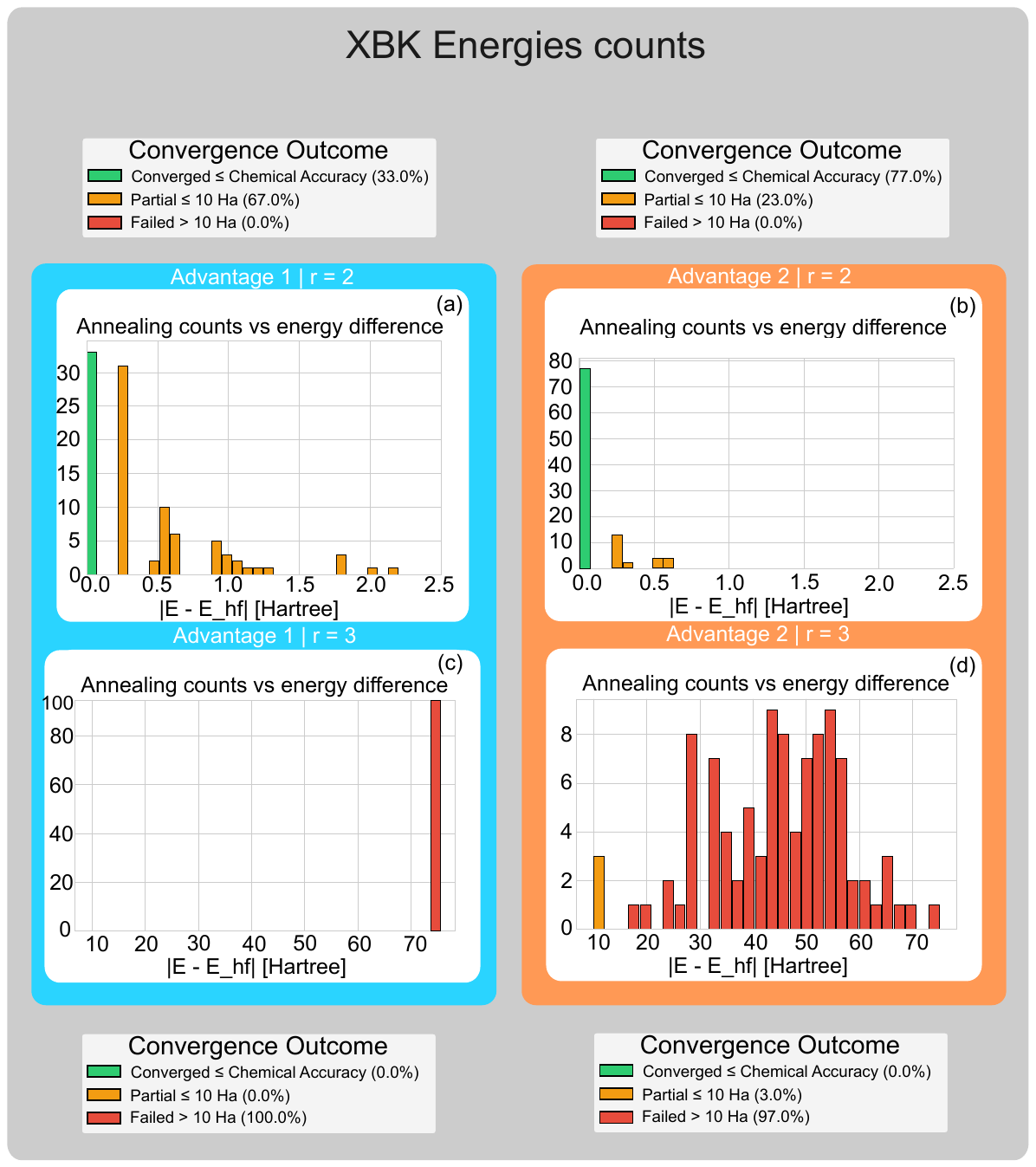} 
    \caption{Annealing Results over 100 runs for the H$_2$O (8,4) active space molecule. Both DWave Advantage 1 (left) and Advantage 2 (right)architectures are tested for the $r$ = 2 and $r$ = 3 parameter, respectively.}
    \label{fig:3}  
\end{figure}

The annealing results shown in Figure~\ref{fig:3} correspond to the H$_2$O molecule in a restricted active space of four orbitals (eight spin-orbitals) and four electrons, evaluated on both the Advantage~1 (AD1) and Advantage~2 (AD2) architectures. To facilitate comparison, the energies reported are expressed as the absolute difference from the Hartree-Fock (HF) reference energy, computed using OpenFermion through classical methods. Tests were carried out for $r=2$, the highest precision level previously benchmarked by Copenhaver \textit{et al.} \cite{copenhaver}, and extended to $r=3$, thereby exceeding the qubit count reported in earlier studies (1.33 times increase in both XBK qubits and logical qubits and about 2.5 times the number of physical qubits). The evaluation of the energy ground state was executed 100 times to show the distribution introduced by the heuristic embedding process.

For $r=2$, AD2 achieved Hartree-Fock (HF) level energies in 77\% of runs (Figure~\ref{fig:3}a), representing a more than doubled improvement over the 33\% success rate of AD1 (Figure~\ref{fig:3}b). Moreover, AD2 consistently produced smaller absolute energy deviations, even in partially converged cases. The method was considered \emph{successful} when the output energy was within chemical accuracy of the HF reference, \emph{partially converged} when the deviation was within 10~Hartree, and \emph{failed} otherwise.

At $r=3$, AD1 failed to converge or produce physically meaningful orbital occupations (Figure~\ref{fig:3}c), consistenly with provous literature of Copenhaver \textit{et al.}. In contrast, AD2 exhibited limited convergence, achieving energies within 10~Hartree of the true value in a few instances (Figure~\ref{fig:3}d). Despite AD2 did not reach full convergence at this precision level, it represents, contrarily to AD1, a qualitative step forward, providing the ground for the refinement discussed later, thanks to physical consistency being the spin configurations corresponding to valid orbital occupations.

\subsection{Connection between the computational resources and the success probability}

In the XBK method, three factors determine the final number of qubits. The first is due to the Hilbert space expansion required to cast the qubit Hamiltonian into an Ising formulation, where the number of qubits increases from \(m\) the size of the Hilbert space to describe the molecule (with \(m = 4\) for the H$_2$O molecule) to \(r m\) -- those already referred to as XBK qubits. The second factor occurs during the quadratisation step, which introduces ancillary qubits to express the Hamiltonian using only quadratic (2-local) terms, here defined as \emph{logical qubits} or \emph{pre-embedding qubits}. Finally, the embedding procedure further enlarges the system by mapping the densely connected logical graph onto the sparse topology of the hardware, resulting in additional \emph{physical qubits} or \emph{post-embedding qubits}.

\begin{figure}  
    \centering
    \includegraphics[width=0.9\linewidth, , trim={0cm 0cm 0cm 0cm}, clip]{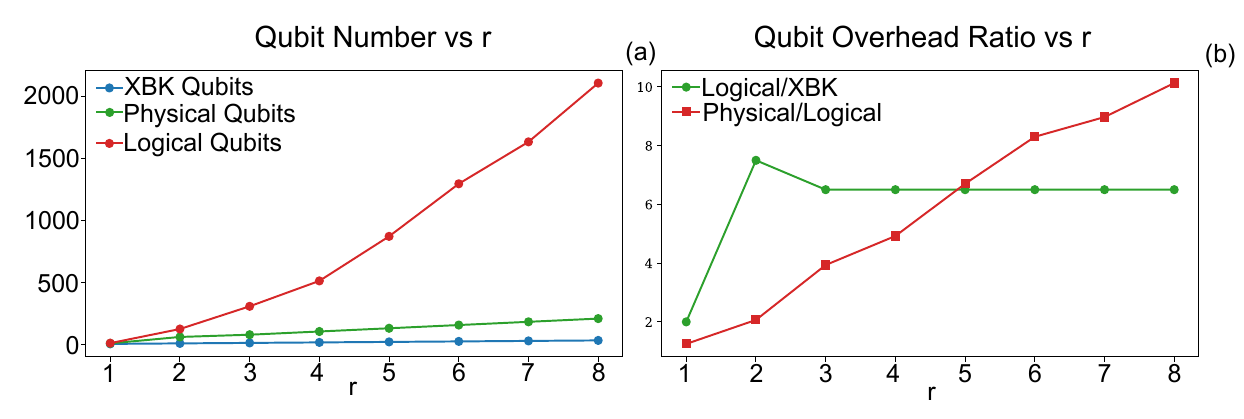} 
    \caption{(a) XBK qubits (blue), Logical qubits (red) and Physical qubits (green) scaling as the $r$ parameter increaases. The qubits required after the embedding are the best case scenario for embeddings found using Advantage 2 connectivity.
    (b) Qubit Overhead Ratio for increasing values of $r$. Increasing the precision has little effect on the number of qubits introduced by quadratisation (green), while it significantly affects the number of physical qubits resulting from the embedding procedure(red).}
    \label{fig:4}  
\end{figure}
For what concerns the latter two factors, Figure~\ref{fig:4} illustrates empirically how the number of pre- and post-embedding qubits scales with the parameter \(r\). While the ratio between XBK and logical qubits remains approximately constant at \(\sim6.5\) as \(r\) increases, the ratio between physical and logical qubits grows steadily, reflecting denser embeddings and longer chains required to compensate for limited hardware connectivity (data from Advantage~2). For \(r=3\), the number of embedded qubits on the hardware is roughly 25 times larger than the corresponding XBK qubits, identifying the principal scalability bottleneck of the method. Although configurations up to \(r=8\) would still occupy only about half of the available QPU space, the average chain lengths greatly surpass the critical threshold of 7-8, beyond which D-Wave hardware becomes unreliable. Classical post-processing can repair broken chains statistically, but the resulting solutions may lose physical validity, ultimately leading to XBK convergence failure.

\begin{figure}  
    \centering
    \includegraphics[width=0.9\linewidth, trim={0cm 0cm 0cm 0cm}, clip]{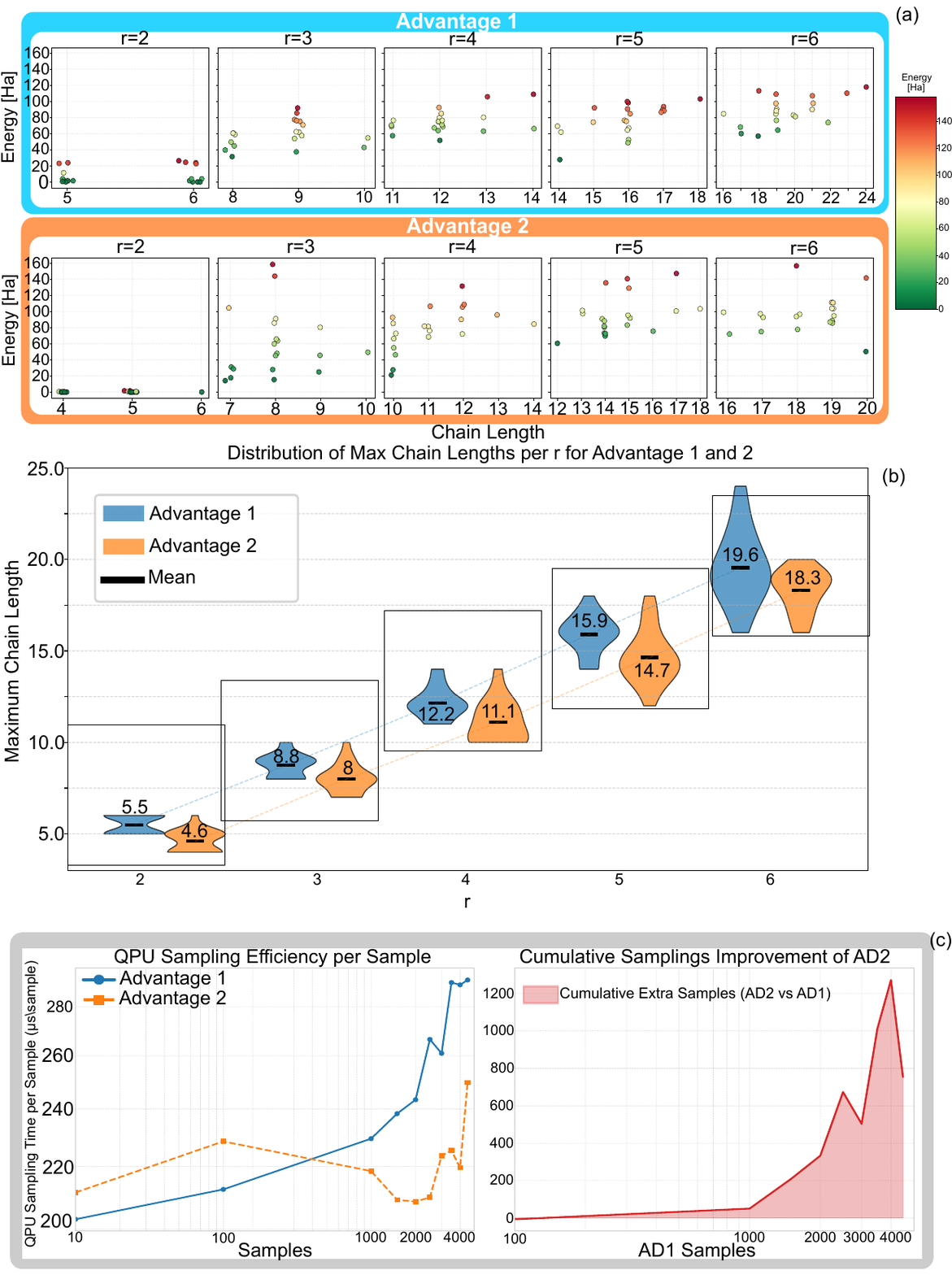} 
    \caption{(a) Comparison of the energy distribution associated with the maximum chain lengths obtained from 20 embedding runs on the Advantage 1 and Advantage 2 architectures, shown for increasing values of r.
(b) Comparison of the best-case maximum chain length distributions for increasing values of r across the Advantage 1 and Advantage 2 architectures.
(c) Comparison between AD1 and AD2 in terms of sampling time per sample as a function of the total number of QPU samples (left). Cumulative number of additional samples that AD2 can perform relative to AD1 (right).}
    \label{fig:5}  
\end{figure}

For a consistent comparison with Advantage~1, the \(P=0, \lambda=0\) Hamiltonian sector is analyzed, as sectors with larger P or $\lambda$ carry too narrow energy gaps to be resolved, as it results from the next Section. Thanks to its denser connectivity, Advantage~2 consistently produces shorter chains, remaining for \(r=3\) within the limit of \text{chain length} $\leq 7\ $ above which the success probability empirically drops, whereas Advantage~1 exceeds such threshold. On average, the \texttt{find\_embedding} algorithm of D-Wave yields chains 1–2 qubits shorter on AD2, enabling physically coherent embeddings for larger problem instances (Figure~\ref{fig:5}b). This reason explains why AD1 statistically fails to provide consistent results at \(r=3\), i.e. the average chain length is too large to preserve coherence, while AD2, when an optimal embedding is found, remains within the reliable operational range.

This fact is experimentally demonstrated in Figure~\ref{fig:5}(a) which compares 20 random embedding runs executed on both AD1 and AD2 for increasing values of \(r\). There, it is shown that AD2 consistently finds natively shorter embeddings, reflecting its denser hardware connectivity. More importantly, a clear correlation emerges between the maximum chain length of an embedding and the likelihood of obtaining a correct solution, indeed problems with shorter chains are statistically more likely to be solved accurately by the quantum annealer. The latter result confirms that chain length represents the primary bottleneck in current quantum annealing approaches to ground-state energy estimation, a limitation fundamentally linked to the need to reformulate the qubit Hamiltonian into its Ising representation.

In terms of execution time, AD2 carries faster QPU sampling, enabling a greater number of iterations within the same runtime (Figure~\ref{fig:5}(c)). Quantitatively, in the 3000–4000 samples range, corresponding to the optimal working regime for H$_2$O at annealing time of 100 microseconds, AD2 performs 16\% to 30\% more samples than AD1.

\subsection{Maximization of the success probability with advanced annealing schedules}

\begin{figure}  
    \centering
    \includegraphics[width=0.9\linewidth,, trim={0cm 0cm 0cm 0cm}, clip]{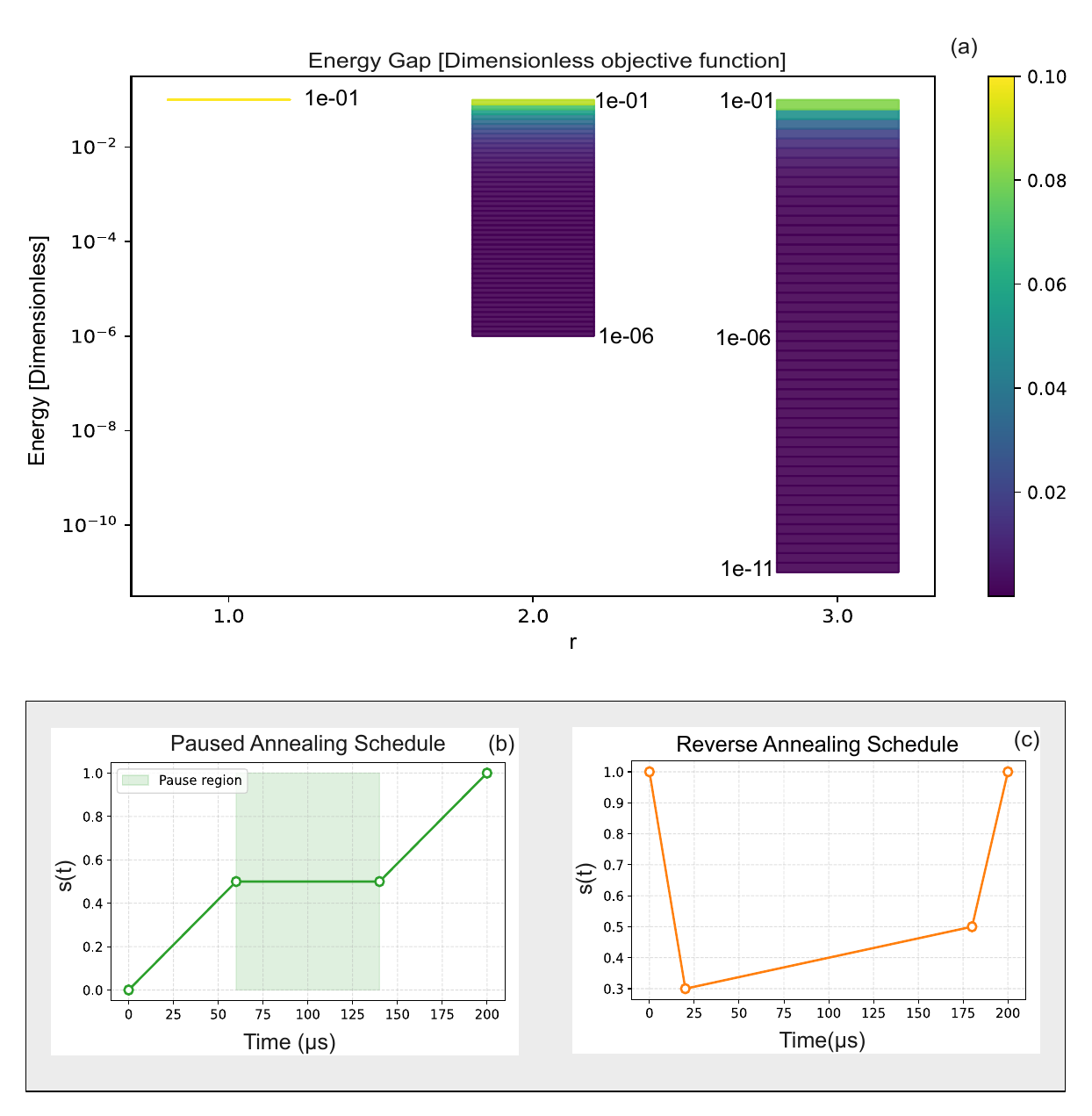} 
    \caption{(a) Energy gap in dimensionless units (the Ising objective function is expressed in internal, unitless values) for increasing values of r. As the method’s precision increases, the problem’s energy landscape demands correspondingly finer energy resolution.
(b) In the paused annealing schedule, the total duration is 200 microseconds. The s parameter is held fixed at 0.5 starting at 80 microseconds, keeping the system in the mixed Hamiltonian for an extended period.
(c) In reverse annealing, the s parameter is rapidly lowered to 0.3 to promote broader state exploration. It is then gradually increased to 0.5 over a 150-microsecond interval, and finally returned to 1 for the final freeze.}
    \label{fig:6}  
\end{figure}

We now turn to advanced annealing schedules for improving ground-state search performance, enabling the achievement of $r=3$ convergence. Prior studies have suggested that tailored schedules can enhance solution quality.\cite{RA} A spectral analysis conducted with Gurobi \cite{gurobi} confirms that, as the $r$ parameter increases, the energy gap of high-P XBK Hamiltonian sectors narrows significantly(Figure~\ref{fig:6} (a)). This explains why, despite AD2 achieves sufficiently short chain lengths to approach Hartree-Fock (HF) accuracy, standard annealing is inadequate and requires refinements.

\begin{figure}  
    \centering
    \includegraphics[width=0.7\linewidth]{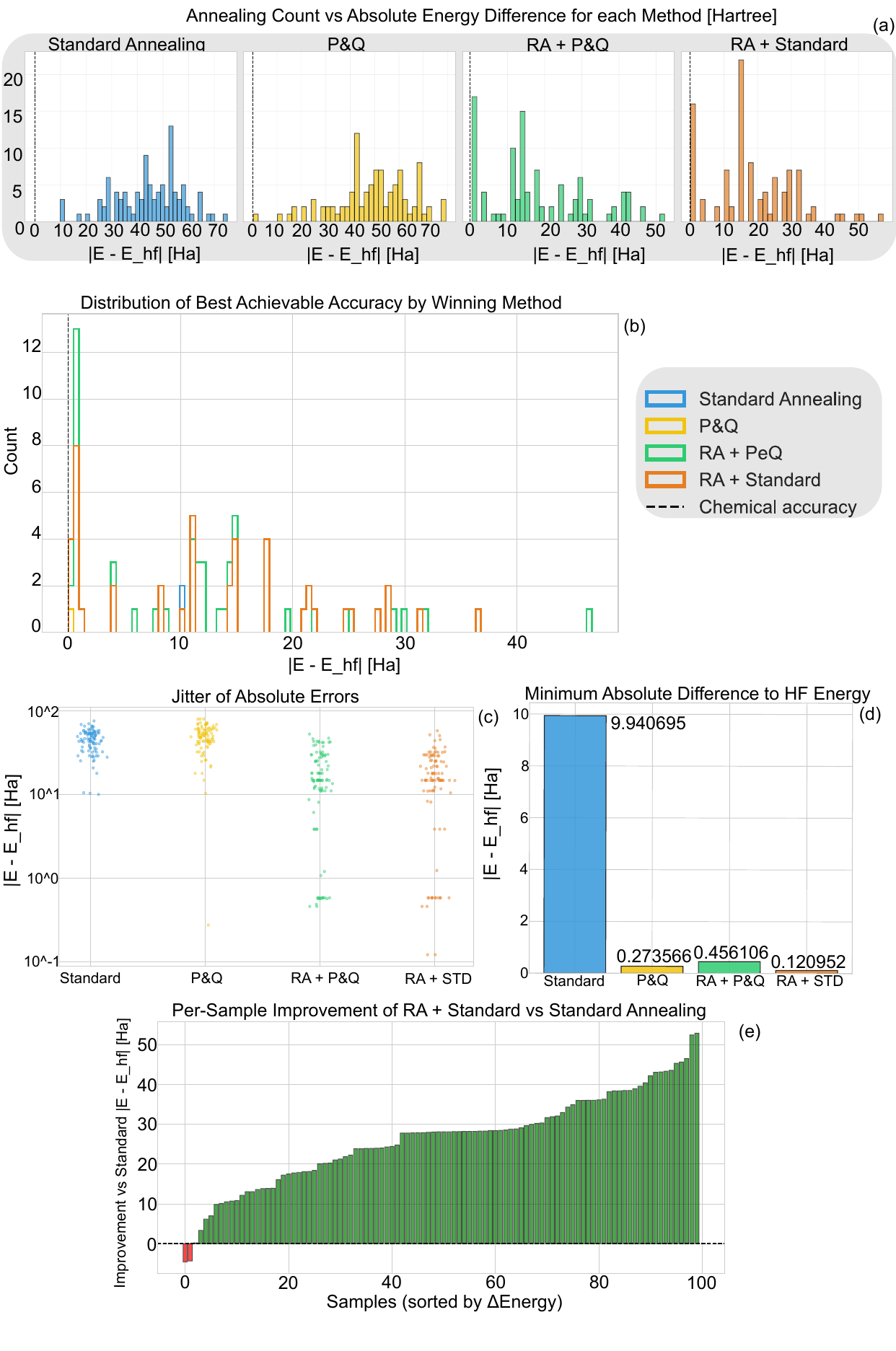} 
    \caption{(a) Histograms of the absolute ground-state energy differences for the four annealing schedules.
    (b)For each of the 100 runs, the annealing schedule that achieved the best (lowest) energy among the four methods is identified and plotted.
    (c)Jitter plot of the absolute differences between the annealing energies and the HF energies for each annealing method.
    (d)Minimum absolute difference between the annealing energies and the HF energies for each annealing method. Reverse annealing, when initialized from samples obtained via standard forward annealing, improves the energy accuracy by nearly two orders of magnitude.
    (e)Energy improvements achieved with reverse annealing, relative to the original energies of the samples obtained via standard forward annealing (which served as the starting states for the reverse-annealing procedure). Reverse annealing improves 98\% of the samples.}
    \label{fig:7}  
\end{figure}

Four methods were tested: (i) standard annealing, modifying only annealing time and number of samples; (ii) paused search; (iii) reverse annealing initialized from standard samples; and (iv) reverse annealing initialized from paused-search samples. For consistency, all tests were performed on the same XBK Hamiltonian sector(Figure~\ref{fig:7}(a)).
Figure~\ref{fig:6}(b) and Figure~\ref{fig:6}(c) respectively show the schedules defined for all Paused and Reverse Annealing schedules. The schedule parameters were selected empirically through iterative testing to identify configurations giving the lowest energies.

The best performance was obtained with reverse annealing from standard samples (Figure~\ref{fig:7}(b)), reaching an energy difference of 0.120~Hartree relative to HF(Figure~\ref{fig:7}(d)). Notably, for complex energy landscapes, reverse annealing refined the standard annealer’s solutions in 98\% of runs(Figure~\ref{fig:7}(e)). Paused search performed comparably to standard annealing, occasionally outperforming reverse annealing in isolated cases(Figure~\ref{fig:7}(c)), but without systematic improvement. Overall, reverse annealing achieved solutions with energy errors two orders of magnitude smaller than the best standard annealing results.

\section{Conclusion}

A systematic benchmarking of the XBK method applied to the H$_2$O molecule within a restricted (8,4) active space using D-Wave’s Advantage~1 and Advantage~2 architectures has been extensively carrier. 
We demonstrated that Advantage~2 consistently achieved Hartree-Fock level energies. Its enhanced qubit connectivity enabled shorter chain lengths and denser yet physically coherent embeddings, extending solvable problem sizes by approximately 130\% relative to previous benchmarks.
Advanced annealing strategies, particularly reverse annealing initialized from standard samples, proved effective in refining solutions, improving accuracy by up to two orders of magnitude, reaching an energy difference of 0.120~Hartree relative to Hartree-Fock.
Overall, these findings demonstrate that improved hardware connectivity, combined with tailored annealing schedules, can significantly enhance the scalability and accuracy of quantum annealing for molecular ground-state estimation. The results show the way for applying XBK-based approaches to larger molecular systems and more complex quantum chemical problems.

\section{Acknowledgements} \par 
This research was supported by the project CQES of the Italian Space Agency (ASI), grant No.2023-46-HH.0.
\printbibliography

\end{document}